\documentclass[showpacs,aps,graphicx,twocolumn]{revtex4}

\usepackage{graphicx}

\begin{document}

%\begin{CJK*}{GBK}{song}

\title{Polarization entanglement purification of nonlocal microwave photons
based on the cross-Kerr effect in circuit QED\footnote{Published in
Phys. Rev. A \textbf{96}, 052330 (2017).}}

\author{Hao Zhang$^{1,2}$, Qian Liu$^{3}$, Xu-Sheng Xu$^{4}$,
Jun Xiong$^{1}$, Ahmed Alsaedi$^{2}$, Tasawar  Hayat$^{2,5}$, and
Fu-Guo Deng$^{1,2}$\footnote{Corresponding author:
fgdeng@bnu.edu.cn}}

\address{$^{1}$Department of Physics, Applied Optics Beijing Area Major Laboratory,
Beijing Normal University, Beijing 100875, China\\
$^{2}$NAAM Research Group, Department of Mathematics, King Abdulaziz
University, Jeddah 21589, Saudi Arabia\\
$^{3}$School of Sciences, Qingdao University of Technology, Qingdao 266033, China\\
$^{4}$State Key Laboratory of Low-Dimensional Quantum Physics and
Department of Physics,
Tsinghua University, Beijing 100084, China\\
$^{5}$Department of Mathematics, Quaid-I-Azam University, Islamabad
44000, Pakistan }

\date{\today}

\begin{abstract}
Microwave photons have become very important qubits in quantum
communication as the first quantum satellite has been launched
successfully. Therefore, it is a necessary and meaningful task for
ensuring the high security and efficiency of microwave-based quantum
communication in practice. Here, we present an original polarization
entanglement purification protocol for nonlocal microwave
photons based on the cross-Kerr effect in circuit quantum
electrodynamics (QED). Our protocol can solve the problem that the
purity of maximally entangled states used for constructing quantum
channels will decrease due to decoherence from environment noise.
This task is accomplished by means of the polarization parity-check
quantum nondemolition (QND) detector, the bit-flipping operation,
and the linear microwave elements. The QND detector is composed of
several cross-Kerr effect systems which can be realized by coupling
two superconducting transmission line resonators to a
superconducting molecule with the $N$-type level structure.
We give the applicable experimental parameters of QND measurement system in circuit QED and analyze the fidelities.
Our protocol has good applications in long-distance quantum communication assisted by microwave photons in
the future, such as satellite quantum communication.
\end{abstract}

 \maketitle

\section{Introduction}\label{sec1}

Quantum entanglement is an indispensable resource for quantum
communication, such as quantum teleportation \cite{CHBennet1993},
quantum dense coding \cite{CHBennet1992,superdense}, quantum key
distribution \cite{AKEkert,bbm92,lixhqkdpra}, quantum secret sharing
\cite{MHillery}, and quantum secure direct communication
\cite{longliupra,FGDeng2003,QSDCWangC,twostepexp}. To
accomplish the quantum communication efficiently,  the two
legitimate remote parties usually use the  photon pairs in maximally
entangled states to construct their communication channel. Due to
the decoherence from the environment in practice, the maximally
entangled state will become a partially entangled state or a mixed
entangled one. This consequence inevitably reduces the efficiency of
the whole communication process. Therefore, some interesting methods
are proposed to improve the efficiency of quantum communication,
such as the error-rejecting coding with decoherence-free subspaces
\cite{DFS1,DFS2,DFS3}, entanglement concentration
\cite{CHBennett1996,YBSheng2008,BCRen012302,Lixhpraecp,HZhangPRA2017},
and entanglement purification
\cite{EPP1,DDeutschPRL1996,JWPannature2001,Simon2002,JWPannature2003,YBShengPRA2008,
EPP2,EPP3,EPP4,dengonestep,EPP5,GYWangPRA2016,CWangECPPRA11,EPPadd1,EPPadd2}.

Entanglement purification is used to transfer a nonlocal mixed
entangled state to a higher purity entangled state. It is a key
technique in quantum repeaters for long-distance quantum
communication to depress the  harmful influence of noise. To date,
some interesting entanglement purification protocols (EPPs) have been
proposed
\cite{EPP1,DDeutschPRL1996,JWPannature2001,Simon2002,JWPannature2003,YBShengPRA2008,
EPP2,EPP3,EPP4,dengonestep,EPP5,GYWangPRA2016,CWangECPPRA11,EPPadd1,EPPadd2}.
For example,  in 1996, Bennett \emph{et al.} \cite{EPP1} proposed an
original EPP for photon pairs in a Werner state
\cite{RFWernerPRA1989} by using two controlled-NOT gates and
single-photon measurements.  In 2001, Pan \emph{et al.}
\cite{JWPannature2001} presented an EPP for a general mixed
entangled state for an ideal entanglement source with simple linear
optical elements. In 2002, Simon and Pan \cite{Simon2002} proposed
an EPP for a nonideal spontaneous parametric down conversion (PDC)
source assisted by using spatial entanglement. In 2003, Pan \emph{et
al.} \cite{JWPannature2003} demonstrated this EPP by using linear
optical elements. In 2008, Sheng \emph{et al.} \cite{YBShengPRA2008}
proposed an efficient polarization EPP for a PDC  source based on the
cross-Kerr effect. In 2010, Sheng and Deng \cite{EPP2} introduced
the original EPP for two-photon systems in a deterministic way.  In
2014, Ren and Deng \cite{EPP5} proposed a two-step hyperentanglement
purification protocol (hyper-EPP) for two-photon four-qubit systems
in nonlocal polarization-spatial hyperentangled Bell states. In
2016, Wang, Liu, and Deng \cite{GYWangPRA2016} presented a universal
method for hyper-EPP in the polarization degree
of freedom and
multiple-longitudinal-momentum degrees of freedom with SWAP gates.

Circuit quantum electrodynamics (QED), which couples the
superconducting qubit to superconducting transmission line
resonators (TLRs), provides a way to study the fundamental
interaction between light and matter \cite{ABlais,AWallraff}. It
holds a big advantage on good scalability for quantum information
processing
\cite{ABlais2,DiCarlo,LongcircuitPRA,Wangsuperconducting,circuitTianlPRL,3q,3q1,RVijaynature,Frederick1}.
Many studies have focused on circuit QED
\cite{AAHouck,MHofheinz,JMajer,DISchuster,BRJohnson,HuaMPRA,Narlaprx2016}. As a
very important and interesting phenomenon, the cross-Kerr effect has been
researched in circuit QED in recent years
\cite{SRebic2009,SKumarPRB2010,YHu,GKirchmair,ICHoi,ETHolland}. For
example, in 2009, Rebi\'{c} \emph{et al.} \cite{SRebic2009} proposed
the giant Kerr nonlinearities at microwave frequencies in circuit
QED. In 2011, Hu \emph{et al.} \cite{YHu} presented a theoretical
scheme to generate the cross-Kerr effect between two TLRs. In 2013,
Hoi \emph{et al.} \cite{ICHoi} observed the giant cross-Kerr effect
for propagating microwaves experimentally induced by an artificial atom. In 2015, Holland \emph{et al.} \cite{ETHolland}
demonstrated the single-photon resolved cross-Kerr effect between
two microwave resonators in experiment. A microwave photon is a very
important qubit for quantum communication because of its low loss and strong
anti-interference during transmission. Due to the decoherence from
environment, the maximally entangled microwave photon state may
become a partially entangled pure state or a mixed one in the
process of transmission and storage. To keep the high
efficiency and fidelity of quantum communication, the  legitimate
parties in quantum communication should make an entanglement
concentration or purification on the partially entangled microwave
photon state or the mixed one, respectively. An original
entanglement concentration protocol has been proposed for microwave
photons \cite{HZhangPRA2017}. To date,  there is no research on
entanglement purification of the nonlocal  entangled states of
microwave-photon pairs. Therefore, the entanglement purification of
microwave-photon states is an extremely important and necessary task
for microwave-based quantum communication. The microwave photon qubit can
be manipulated effectively \cite{Narlaprx2016,JMHao,DRSolli1}. For example, Narla \emph{et al.} \cite{Narlaprx2016} realized the basic microwave beam splitter which plays a very important role for microwave homodyne detection and used it to generate the robust concurrent remote entanglement between two superconducting qubits. The polarization can be manipulated by adjusting the material parameters \cite{JMHao,DRSolli1}.

In this paper, we propose a physically feasible polarization
EPP on the nonlocal entangled microwave photons in circuit QED.
By using our EPP, the parties can effectively purify the mixed
entangled states induced by  the decoherence from environment noise
in microwave-based quantum communication. This task is achieved with
the polarization parity-check QND measurements on microwave-photon
pairs, the bit-flipping operations, and the linear microwave
elements. The parity-check quantum nondemolition (QND) detector is composed of two
cross-Kerr systems for microwave photons and is a crucial part to
implement the polarization entanglement purification. We give the
applicable experimental parameters of a QND measurement system and
analyze the fidelities. The protocol has some good applications in
nonlocal microwave-based quantum communication, such as satellite
quantum communication.

This article is organized as follows:  We first review the
cross-Kerr effect in circuit QED in Sec.~\ref{sec21} and then
describe the process for the QND measurement on two cascade TLRs in
Sec.~\ref{sec22}. We present an EPP for microwave-photon pairs in
Sec.~\ref{sec31} and perform the EPP for polarization-spatial
entangled microwave-photon pairs in Sec.~\ref{sec32}. In
Sec.~\ref{sec33}, we design the reasonable parameters for QND
measurement systems and analyze the fidelities. A
summary is given in Sec.~\ref{sec4}.

\begin{figure}[!ht]%[tpb]
\begin{center}
\includegraphics[width=8.0cm,angle=0]{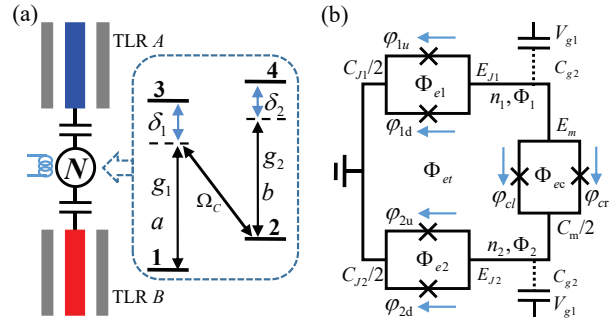}
\caption{(a) Schematic diagram of the cross-Kerr effect induced by
coupling TLR \emph{A} (top, blue) and \emph{B} (bottom, red) to a
superconducting molecule (middle, circle with \emph{N}). The
molecule can be controlled by external coils (left coils). The
\emph{N}-type level structure of the artificial molecule is shown in
the right dashed line box. (b) The  structure of superconducting
quantum circuit for the molecule \cite{YHu}.} \label{fig1}
\end{center}
\end{figure}

\section{The QND measurement system in circuit QED} \label{basic}

\subsection{Cross-Kerr effect between two TLRs}
\label{sec21}

The schematic diagram for realizing the cross-Kerr effect between
two TLRs is shown in Fig.~\ref{fig1}. The cross-Kerr effect can be
realized by coupling two TLRs to a four level \emph{N}-type
superconducting molecule as shown in Fig.~\ref{fig1}(a). The level
structure is depicted in the dashed line box. TLR \emph{A} and TLR
\emph{B} are coupled to the levels $1-3$ and $2-4$, respectively.
The transition between the levels $2$ and $3$ is driven by a
classical pump laser with the strength $\Omega_{c}$. In the interaction picture, the Hamiltonian
of the whole interaction system is given by \cite{YHu} (with
$\hbar=1$)
\begin{eqnarray}        \label{eq1}
\hat{H}&=&\delta_{1}\hat{\sigma}_{33}+\delta_{2}\hat{\sigma}_{44}
+i\text{g}_{1}\left(\hat{\sigma}_{13}\hat{a}^{\dag}-\hat{\sigma}_{31}\hat{a}\right)\nonumber\\
&&+i\text{g}_{2}\left(\hat{\sigma}_{24}\hat{b}^{\dag}-\hat{\sigma}_{42}\hat{b}\right)
+i\Omega_{c}\left(\hat{\sigma}_{23}-\hat{\sigma}_{32}\right),\;\;\;\;\;\;
\end{eqnarray}
where the detunings are $\delta_{1}=E_{31}-\omega_{1}$ and
$\delta_{2}=E_{42}-\omega_{2}$. $\omega_{1}$ and $\omega_{2}$ are
the frequencies of the TLRs \emph{A} and \emph{B}, respectively.
$\hat{\sigma}_{ij}=|i\rangle\langle j|$ is the transition operator
from the states $|j\rangle$ to $|i\rangle$. $\hat{a}$
$(\hat{a}^{\dag})$ and $\hat{b}$ $(\hat{b}^{\dag})$ are the
annihilation (creation) operators for the modes of TLRs \emph{A} and \emph{B}, respectively. $\text{g}_{1}$ and $\text{g}_{2}$ are the
coupling strengths for corresponding interactions between TLRs and
levels. Under the conditions that
$\mid\text{g}_{1}/\Omega_{c}\mid^{2}\ll 1$ and
$\mid\text{g}_{2}\mid\ll \mid\delta_{2}\mid$ \cite{AImamoglu1997},
one can adiabatically eliminate the atomic degrees of freedom and
obtain the effective cross-Kerr interaction Hamiltonian \cite{YHu}
\begin{eqnarray}        \label{eq2}
\hat{H}_{K}=\chi\hat{a}^{\dag}\hat{a}\hat{b}^{\dag}\hat{b},
\end{eqnarray}
where
$\chi=-\text{g}_{1}^{2}\text{g}_{2}^{2}/(\delta_{2}\Omega^{2}_{c})$
is the cross-Kerr coefficient.

The  molecule with an \emph{N}-type level structure can be
constructed in the superconducting circuit described in
Fig.~\ref{fig1}(b). The two loops (bottom and top) are two transmon
qubits \cite{JKoch}. The right loop is a superconducting quantum
interference device (SQUID) \cite{JSiewert} which is used to connect
two transmon qubits. Each loop is composed of two identical
Josephson junctions labeled with crosses. $C_{j}/2$ $(j=m,1,2)$ and
$E_{Ji}$ $(i=c,1,2)$ represent the capacitance and energy of the
Josephson junctions, respectively. The gate voltages labeled with
$V_{\text{g}1}$ and $V_{\text{g}2}$ bias the corresponding transmons
via the gate capacitors $C_{\text{g}1}$ and $C_{\text{g}2}$,
respectively. $\Phi_{e1}$, $\Phi_{e2}$, $\Phi_{ec}$, and $\Phi_{et}$
are external fluxes. $\varphi_{cr}$, $\varphi_{cl}$, $\varphi_{1u}$,
$\varphi_{1d}$, $\varphi_{2u}$, and $\varphi_{2d}$ are the
gauge-invariant phases across the Josephson junctions. By using the
two-level language in the region $E_{J}\gg E_{c}$, one can obtain
the $N$-type level form \cite{YHu,JKoch} shown in
Fig.~\ref{fig1}(a). The eigenstates and the corresponding
eigenvalues are $|i\rangle$ and $E_{i}$ $(i=1,2,3,4)$, respectively.

\begin{figure}[!ht]%[tpb]
\begin{center}
\includegraphics[width=7.0cm,angle=0]{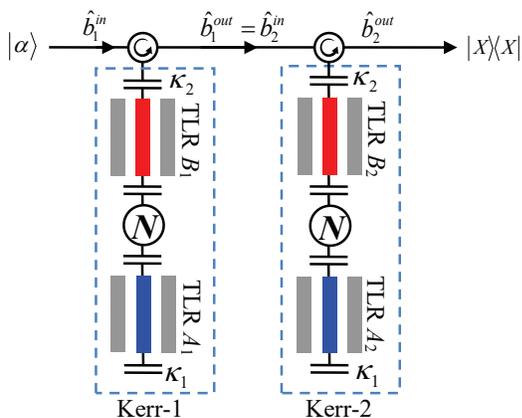}
\caption{ Schematic diagram of QND measurement of the total
photon number of the two TLRs labeled with $A$ (i.e., $A_1$ and
$A_2$). The two TLRs labeled with $B$  (i.e., $B_1$ and $B_2$) with
the same decay rate $\kappa_{2}$ are the readout resonators and all
the TLRs labeled with \emph{A} with the decay rate $\kappa_{1}$ are
the storage resonators. The circle with \emph{N} stands for a
superconducting molecule with the \emph{N}-type level structure. The
direction of the arrow represents the spread direction of the probe
light. The elements labeled with a circular arrow in a big circle
are circulators. $|\alpha\rangle$ represents the probe light.
$|X\rangle\langle X|$ represents the homodyne measurement on the
coherent state of the probe light. } \label{fig2}
\end{center}
\end{figure}

\subsection{The Quantum nondemolition measurement on total photon number of transmission-line resonators based on cross-Kerr effect}
\label{sec22}

The total photon number of TLRs can be measured with QND by means of the
cross-Kerr effect. The detailed schematic diagram is shown in
Fig.~\ref{fig2}. All the TLRs in the top and the bottom are the
readout and storage resonators, respectively. The Homedyne detection
has been used for microwave in circuit QED experiment
\cite{RVijaynature}. The probe light in the coherent state
$|\alpha\rangle$ is input from the left and measured via an $X$
homodyne measurement on the right. Here, we use the input-output
relationship to explain the whole process. When the probe light is
resonant with readout resonators, the Heisenberg-Langevin equations
for each cross-Kerr media in the probe light path are given by
\begin{eqnarray}        \label{eqlangevin}
\dot{\hat{b}_{k}}=-i\chi_{k}\,
\hat{n}_{k}\hat{b}_{k}-\frac{\kappa_{2}}{2}\hat{b}_{k}-\sqrt{\kappa_{2}}\,\hat{b}^{in}_{k},
\end{eqnarray}
where $k=1$, $2$ and $\hat{n}_{k}=\hat{a}_{k}^{\dag}\hat{a}_{k}$
represents the photon number operator of the $k$-th storage
resonator. We assume that all TLRs labeled with \emph{A} and
\emph{B} have the same decay rates $\kappa_{1}$ and $\kappa_{2}$,
respectively.

Now, we consider the situation that the decay rate of the readout
resonator $\kappa_{2}\gg \chi_{k}\langle \hat{n}_{k}\rangle$. One
can make $\dot{\hat{b}_{k}}=0$ in Eq.~(\ref{eqlangevin}). Combining
with the standard cavity input-output relationship
$\hat{b}_{out}=\hat{b}_{in}+\sqrt{\kappa_{2}}\,\hat{b}$
\cite{DFWalls,Kevin}, where $\hat{b}_{in}$ and $\hat{b}^{\dag}_{in}$
satisfy the standard commutation relations
$[\hat{b}_{in}(t),\hat{b}^{\dag}_{in}(t^{\prime})]=\delta(t-t^{\prime})$,
one can obtain the reflection coefficients which are expressed as
\begin{eqnarray}        \label{eq4}
r_{k}(\hat{n}_{k})=\frac{\hat{b}^{out}_{k}}{\hat{b}^{in}_{k}}
=\frac{i\chi_{k} \hat{n}_{k}-\frac{\kappa_{2}}{2}}{i\chi_{k}\hat{n}_{k}+\frac{\kappa_{2}}{2}}.
\end{eqnarray}

Our goal is to make a QND measurement on the total photon number in
two storage resonators (the two TLRs $A_1$ and $A_2$ in the bottom
of Fig.~\ref{fig2}). For this task, a probe light in the coherent
state
$|\alpha\rangle^{in}$
is input from the left, and let us assume that there are $n_{1}$
photons in TLR $A_{1}$. When the probe light leaves TLR $B_{1}$, the
state of Kerr-1 becomes
\begin{eqnarray}        \label{eq5}
|\psi\rangle^{out}_{K1}=|n_{1}\rangle |e^{i\theta_{n1}}\alpha\rangle_{1}^{out}.
\end{eqnarray}
Here $\theta_{n1}= $ arg$\left[r_{1}(n_{1})\right]$ and
$|n_{1}\rangle$ is a Fock state.
One can make an $X$ homodyne measurement to infer the photon number
in TLR $A_{1}$ because the phase shift depends on the photon number
$n_{1}$. When the probe light passes through TLR $B_{2}$, two Kerr
media become a cascaded system. Therefore, we set
$\hat{b}^{in}_{2}=\hat{b}^{out}_{1}$ because the input field of resonator
$B_{2}$ is the output field of resonator $B_{1}$. The input-output
relationship of this cascaded system is
$\hat{b}^{out}_{2}=r_{2}(n_{2}) r_{1}(n_{1})\hat{b}^{in}_{1}$.
Let us assume that the photon number in TLR $A_{2}$ is $n_{2}$.
After the probe light leaves resonator $B_{2}$, its state is given
by
\begin{eqnarray}        \label{eq6}
|\alpha\rangle_{2}^{out}\!=\!|e^{i\theta_{n1+n2}}\alpha\rangle_{2}^{out},\;\;
\end{eqnarray}
where
\begin{eqnarray}        \label{eq7}
\theta_{n1+n2}=\theta_{n1}+\theta_{n2}=\text{arg}\left[r_{1}(n_{1})
\cdot r_{2}(n_{2})\right],
\end{eqnarray}
where $\theta_{n2}=\text{arg}\left[r_{2}(n_{2})\right]$.

To make an effective homodyne detection, we detect the
position quadrature $X$ of the coherent state. The wave function in the
coherent state is given by \cite{KNemotoprl2004,SDBarrettpra}
\begin{eqnarray}        \label{homodyne}
\langle X|\alpha e^{i\theta}\rangle=f(X,\alpha \cos\theta)e^{i\Phi(X)},
\end{eqnarray}
where the functions are given by
\begin{eqnarray}        \label{homodynefunction}
&f(X,y)&=\frac{1}{\sqrt[4]{2\pi}}\exp[-\frac{1}{4}(x-2y)^{2}]\nonumber \\ &\Phi(X)&=\alpha\sin\theta(x-2\alpha\cos\theta)\text{mod}(2\pi).
\end{eqnarray}
Therefore, for states $|\alpha e^{i\theta_{1}}\rangle$ and $|\alpha e^{i\theta_{2}}\rangle$, the midpoint and distance between the peaks of corresponding functions $f(X,\alpha\cos\theta_{1})$ and $f(X,\alpha\cos\theta_{2})$ are $X_{m}=\alpha(\cos\theta_{1}+\cos\theta_{2})$ and $X_{d}=2\alpha(\cos\theta_{1}-\cos\theta_{2})$, respectively. According to the result of position, one can distinguish the different phases. The error probability is given by \cite{SDBarrettpra}
\begin{eqnarray}        \label{erfc}
P_{\text{error}}=\frac{1}{2}\text{erfc}\bigg[\frac{X_{d}}{2\sqrt{2}}\bigg],
\end{eqnarray}
where $\text{erfc}(x)$ is the complementary error function.

When we only consider the maximal total photon number of two, all the different Fock states and corresponding phase shifts are shown in Table~\ref{Tab1}.

\begin{table}
  \begin{center}
\caption{The corresponding relation between the states of the signal
light and the phase shifts.}
\begin{tabular}{lccccccccccccccccccccccccccccccccc}
\hline\hline $|\hat{a}_{1}\rangle\otimes|\hat{a}_{2}\rangle$
&&&&&&&&&&&&&&&&&&&&&&&&&&&&&&&&& Total phase shift  \\ \hline
$|0\rangle\otimes|0\rangle$ &&&&&&&&&&&&&&&&&&&&&&&&&&&&&&&&& $\text{arg}
\left[r_{1}(0)\cdot r_{2}(0)\right]$   \\ %\hline
$|1\rangle\otimes|0\rangle$ &&&&&&&&&&&&&&&&&&&&&&&&&&&&&&&&& $\text{arg}
\left[r_{1}(1)\cdot r_{2}(0)\right]$  \\ %\hline
$|2\rangle\otimes|0\rangle$ &&&&&&&&&&&&&&&&&&&&&&&&&&&&&&&&& $\text{arg}
\left[r_{1}(2)\cdot r_{2}(0)\right]$  \\ %\hline\hline
$|0\rangle\otimes|1\rangle$ &&&&&&&&&&&&&&&&&&&&&&&&&&&&&&&&& $\text{arg}
\left[r_{1}(0)\cdot r_{2}(1)\right]$   \\ %\hline
$|1\rangle\otimes|1\rangle$ &&&&&&&&&&&&&&&&&&&&&&&&&&&&&&&&& $\text{arg}
\left[r_{1}(1)\cdot r_{2}(1)\right]$  \\ %\hline
$|0\rangle\otimes|2\rangle$ &&&&&&&&&&&&&&&&&&&&&&&&&&&&&&&&& $\text{arg}
\left[r_{1}(0)\cdot r_{2}(2)\right]$ \\
\hline\hline
\end{tabular}\label{Tab1}
\end{center}
\end{table}

\begin{figure*}[!ht]%[tpb]
\begin{center}
\includegraphics[width=14.4cm,angle=0]{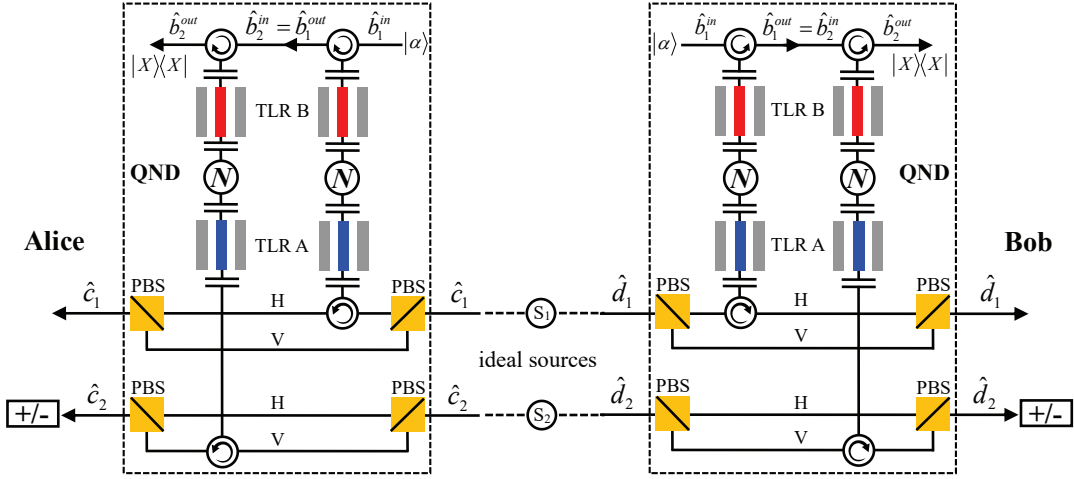}
\caption{ Schematic diagram for the entanglement purification on two
microwave-photon pairs. $S_{1}$ and $S_{2}$ are the two identical
ideal entanglement sources for microwave-photon pairs. Two dashed
boxes are two same-polarization parity-check QND detectors. PBS
represents a polarizing beam splitter for microwave photons. The
circles with a circular arrow stands for circulators. The QND
measurement is given in Fig.~\ref{fig2}. Two rectangular boxes
labeled with $+/-$ signs are two measurements with the two diagonal
bases $\{|\pm\rangle=\frac{1}{\sqrt{2}}(|H\rangle\pm|V\rangle)\}$.}
\label{figepid}
\end{center}
\end{figure*}

\section{Entanglement purification of bit-flipping errors for microwave photons}
\label{sec3}

\subsection{Entanglement purification protocol for microwave-photon pairs}
\label{sec31}

Let us assume that the nonlocal microwave-photon pairs in quantum
communication are in the mixed state $\hat{\rho}_{cd}$ described by
\begin{eqnarray}        \label{rhocd}
\hat{\rho}_{cd}=f|\Phi^{\dag}\rangle_{cd}\langle \Phi^{\dag}|
+(1-f)|\Psi^{\dag}\rangle_{cd}\langle \Psi^{\dag}|,
\end{eqnarray}
where
\begin{eqnarray}        \label{Bellpolarization}
|\Phi^{\dag}\rangle_{cd}&=&\frac{1}{\sqrt{2}}(|H\rangle_{c}|H\rangle_{d}
+|V\rangle_{c}|V\rangle_{d}),\nonumber\\
|\Psi^{\dag}\rangle_{cd}&=&\frac{1}{\sqrt{2}}(|H\rangle_{c}|V\rangle_{d}
+|V\rangle_{c}|H\rangle_{d}).
\end{eqnarray}
$H$ and $V$ represent the horizontal and the vertical polarizations
of microwave photons, respectively. The symbol $f$ with the
relationship $f=\langle
\Phi^{\dag}|\hat{\rho}_{cd}|\Phi^{\dag}\rangle$ is the fidelity of
the state $ |\Phi^{\dag}\rangle$ ($f>\frac{1}{2}$). In this way, the
state of the system composed of two microwave-photon pairs is just
the mixture of four states. They are
$|\Phi^{\dag}\rangle_{c1d1}|\Phi^{\dag}\rangle_{c2d2}$ with a
probability of $f^{2}$,
$|\Phi^{\dag}\rangle_{c1d1}|\Psi^{\dag}\rangle_{c2d2}$ and
$|\Psi^{\dag}\rangle_{c1d1}|\Phi^{\dag}\rangle_{c2d2}$ with the same
probability of $(1-f)f$, and
$|\Psi^{\dag}\rangle_{c1d1}|\Psi^{\dag}\rangle_{c2d2}$ with a
probability of $(1-f)^{2}$.

The principle of our EPP for the polarization entanglement of
nonlocal microwave-photon pairs from two identical ideal
entanglement sources is shown in Fig.~\ref{figepid}. Here, we choose
two same cross-Kerr systems, i.e., $\chi_{1}=\chi_{2}$, to accomplish
the QND measurement process for parity check. We will discuss the
physical implementation in Sec.~\ref{sec33}. The microwave
polarizing beam splitter (PBS) shown in Fig.~\ref{figepid} can pass
 the  photons in the state $|H\rangle$ and reflect the
photons in the state $|V\rangle$. Therefore, in the QND part of this
protocol, we can change $|H\rangle$ and $|V\rangle$ to $|1\rangle$
and $|0\rangle$ for $\hat{c}_{1}\hat{d}_{1}$, respectively. For
$\hat{c}_{2}\hat{d}_{2}$, $|H\rangle$ and $|V\rangle$ can be
represented by $|0\rangle$ and $|1\rangle$, respectively. The
different polarization states and corresponding phase shifts are
rewritten in Table~\ref{Tab2}. The two QND measurement detectors are
identical and the two parties in quantum communication Alice and Bob
hold $\hat{c}$ and $\hat{d}$, respectively. We don't consider
the phase difference of microwave photon after it leaves the
storage resonator in our scheme, because we just design
the principle here. The possible phase difference can be
compensated in practice.

\begin{table}
\begin{center}
\caption{ The corresponding relation between the states of the
signal light and the phase shifts by using the same
cross-Kerr media in each QND detector.}
\begin{tabular}{lcccccccccccccccccccccccc}
\hline\hline
$\hat{c}_{1}\hat{c}_{2}/\hat{d}_{1}\hat{d}_{2}$($|\hat{a}_{1}\rangle|\hat{a}_{2}\rangle$)
&&&&&&&&&&&&&&&&&&&&&&&& Total phase shift  \\ \hline
$|V\rangle|V\rangle\rightarrow(|0\rangle|1\rangle)$ &&&&&&&&&&&&&&&&&&&&&&&& $\theta_{1}$  \\ %\hline
$|H\rangle|H\rangle\rightarrow(|1\rangle|0\rangle)$ &&&&&&&&&&&&&&&&&&&&&&&& $\theta_{1}$   \\
$|H\rangle|V\rangle\rightarrow(|1\rangle|1\rangle)$ &&&&&&&&&&&&&&&&&&&&&&&& $\theta_{2}$ \\ %\hline
$|V\rangle|H\rangle\rightarrow(|0\rangle|0\rangle)$ &&&&&&&&&&&&&&&&&&&&&&&& $\theta_{0}$ \\ \hline\hline
\end{tabular}\label{Tab2}
\end{center}
\end{table}

When the microwave-photon pairs in the state
$|\Phi^{\dag}\rangle_{c1d1}|\Phi^{\dag}\rangle_{c2d2}$ pass through
the parity-check QND detectors, the state of the composite system
composed of the two microwave-photon pairs ($\hat{c}_1\hat{d}_1$ and
$\hat{c}_2\hat{d}_2$) and the two probe lights ($\hat{c}$ and
$\hat{d}$) becomes
\begin{eqnarray}\label{eqnoerror}
\Longrightarrow\!\!\!\!&&\frac{1}{2}\{(|H\rangle_{c1}|H\rangle_{d1}
|H\rangle_{c2}|H\rangle_{d2}\nonumber\\
&&+|V\rangle_{c1}|V\rangle_{d1}|V\rangle_{c2}|V\rangle_{d2})
|\alpha e^{i\theta_{1}}\rangle_{c}|\alpha e^{i\theta_{1}}\rangle_{d}\nonumber\\
&&+|H\rangle_{c1}|H\rangle_{d1}|V\rangle_{c2}|V\rangle_{d2}
|\alpha e^{i\theta_{2}}\rangle_{c}|\alpha e^{i\theta_{2}}\rangle_{d}\nonumber\\
&&+|V\rangle_{c1}|V\rangle_{d1}|H\rangle_{c2}|H\rangle_{d2} |\alpha
e^{i\theta_{0}}\rangle_{c}|\alpha e^{i\theta_{0}}\rangle_{d}\}.\;\;\;\;
\end{eqnarray}
When Alice and Bob obtain a phase shift with $\theta_{1}$ on their
coherent states after the homodyne detections, the state will
collapse to
$(|H\rangle_{c1}|H\rangle_{d1}|H\rangle_{c2}|H\rangle_{d2}
+|V\rangle_{c1}|V\rangle_{d1}|V\rangle_{c2}|V\rangle_{d2})$. For the
last two terms, there are two situations. If
$\theta_{2}\!=\!\theta_{0}\!+\!2\pi$, the last two terms have the
same phase shifts. At this point, both Alice and Bob obtain the
phase shift $\theta_{0}$ on their coherent states, the state becomes
$(|H\rangle_{c1}|H\rangle_{d1}|V\rangle_{c2}|V\rangle_{d2}
+|V\rangle_{c1}|V\rangle_{d1}|H\rangle_{c2}|H\rangle_{d2})$.
Subsequently, Alice and Bob can perform a bit-flipping operation
$\hat{\sigma}_{x}=|H\rangle\langle V|+|V\rangle\langle H|$ on
$\hat{c}_{1}$ and $\hat{d}_{1}$, respectively, and then they can
obtain the state
$(|H\rangle_{c1}|H\rangle_{d1}|H\rangle_{c2}|H\rangle_{d2}
+|V\rangle_{c1}|V\rangle_{d1}|V\rangle_{c2}|V\rangle_{d2})$. If the
$\theta_{2}\neq\theta_{0}+2\pi$, Alice and Bob will obtain the
different results, they will make no operation.  To get the
state $|\Phi^{\dag}\rangle_{cd}$, Alice and Bob make a measurement
with the diagonal basis
$\{|\pm\rangle=\frac{1}{\sqrt{2}}(|H\rangle\pm|V\rangle)\}$ on
$\hat{c}_{2}$ and $\hat{d}_{2}$, respectively. When both their
results are $|+\rangle$ or $|-\rangle$, the state of the
microwave-photon pair $\hat{c}_{1}\hat{d}_{1}$ becomes
$|\Phi^{\dag}\rangle_{cd}$. Otherwise, they should make the
operation $\hat{\sigma}_{z}=|H\rangle\langle H|-|V\rangle\langle V|$
on the microwave photon $\hat{c}_{1}$ to obtain the state
$|\Phi^{\dag}\rangle_{c_1d_1}$.

After the QND measurement process, the state of the system
$|\Phi^{\dag}\rangle_{c1d1}|\Psi^{\dag}\rangle_{c2d2}|\alpha\rangle_{c}|\alpha\rangle_{d}$
becomes
\begin{eqnarray}\label{eq11}
\Longrightarrow\!\!\!\!&&\frac{1}{2}\{|H\rangle_{c1}|H\rangle_{d1}|V\rangle_{c2}
|H\rangle_{d2}|\alpha e^{i\theta_{2}}\rangle_{c}|\alpha e^{i\theta_{1}}\rangle_{d}\nonumber\\
&&+|V\rangle_{c1}|V\rangle_{d1}|H\rangle_{c2}|V\rangle_{d2}|
\alpha e^{i\theta_{0}}\rangle_{c}|\alpha e^{i\theta_{1}}\rangle_{d}\nonumber\\
&&+|H\rangle_{c1}|H\rangle_{d1}|H\rangle_{c2}|V\rangle_{d2}|
\alpha e^{i\theta_{1}}\rangle_{c}|\alpha e^{i\theta_{2}}\rangle_{d}\nonumber\\
&&+|V\rangle_{c1}|V\rangle_{d1}|V\rangle_{c2}|H\rangle_{d2}| \alpha e^{i\theta_{1}}
\rangle_{c}|\alpha e^{i\theta_{0}}\rangle_{d}\}.\;\;\;\;\;\;
\end{eqnarray}
Another state
$|\Psi^{\dag}\rangle_{c1d1}|\Phi^{\dag}\rangle_{c2d2}|\alpha\rangle_{c}|\alpha\rangle_{d}$
is evolved to
\begin{eqnarray}\label{eq12}
\Longrightarrow\!\!\!\!&&\frac{1}{2}\{|V\rangle_{c1}|H\rangle_{d1}|H\rangle_{c2}
|H\rangle_{d2}|\alpha e^{i\theta_{0}}\rangle_{c}|\alpha e^{i\theta_{1}}\rangle_{d}\nonumber\\
&&+|H\rangle_{c1}|V\rangle_{d1}|V\rangle_{c2}|V\rangle_{d2}|\alpha
e^{i\theta_{2}}\rangle_{c}
|\alpha e^{i\theta_{1}}\rangle_{d}\nonumber\\
&&+|V\rangle_{c1}|H\rangle_{d1}|V\rangle_{c2}|V\rangle_{d2}|\alpha e^{i\theta_{1}}
\rangle_{c}
|\alpha e^{i\theta_{2}}\rangle_{d}\nonumber\\
&&+|H\rangle_{c1}|V\rangle_{d1}|H\rangle_{c2}|H\rangle_{d2}|\alpha e^{i\theta_{1}}
\rangle_{c} |\alpha e^{i\theta_{0}}\rangle_{d}\}.\;\;\;\;\;\;
\end{eqnarray}
From these two results, with the situation
$\theta_{2}=\theta_{0}+2\pi$, one can see that if Alice and Bob
obtain the phase shifts $\theta_{0}$ and $\theta_{1}$ with a homodyne measurement on
their probe lights $\hat{c}$ and $\hat{d}$, the state of the two
microwave-photon pairs $\hat{c}_1\hat{d}_1\hat{c}_2\hat{d}_2$
becomes $(|H\rangle_{c1}|H\rangle_{d1}|V\rangle_{c2}|H\rangle_{d2}
+|V\rangle_{c1}|V\rangle_{d1}|H\rangle_{c2}|V\rangle_{d2})$ or
$(|V\rangle_{c1}|H\rangle_{d1}|H\rangle_{c2}|H\rangle_{d2}
+|H\rangle_{c1}|V\rangle_{d1}|V\rangle_{c2}|V\rangle_{d2})$. As
Alice and Bob cannot determine on which pair a bit-flipping error
occurs, they discard both photon pairs in these two situations. With the
situation $\theta_{2}\neq\theta_{0}+2\pi$, Alice and Bob get the
different results for all the terms. They should also discard all
these situations.

After the microwave-photon pairs pass through the QND detectors,
the state of the system
$|\Psi^{\dag}\rangle_{c1d1}|\Psi^{\dag}\rangle_{c2d2}|\alpha\rangle_{c}|\alpha\rangle_{d}$
turns to
\begin{eqnarray}\label{eq13}
\Longrightarrow\!\!\!\!&&\frac{1}{2}\{(|V\rangle_{c1}|H\rangle_{d1}
|V\rangle_{c2}|H\rangle_{d2}\nonumber\\
&&+|H\rangle_{c1}|V\rangle_{d1}|H\rangle_{c2}|V\rangle_{d2})
|\alpha e^{i\theta_{1}}\rangle_{c}|\alpha e^{i\theta_{1}}\rangle_{d}\nonumber\\
&&+|V\rangle_{c1}|H\rangle_{d1}|H\rangle_{c2}|V\rangle_{d2}
|\alpha e^{i\theta_{0}}\rangle_{c}|\alpha e^{i\theta_{2}}\rangle_{d}\nonumber\\
&&+|H\rangle_{c1}|V\rangle_{d1}|V\rangle_{c2}|H\rangle_{d2} |\alpha
e^{i\theta_{2}}\rangle_{c}|\alpha
e^{i\theta_{0}}\rangle_{d}\}.\;\;\;\;\;\;\;\;
\end{eqnarray}
The result is similar to that in the situation with no bit-flipping
error expressed in Eq.~(\ref{eqnoerror}). Due to the
indistinguishability with the situation with no bit-flipping errors,
Alice and Bob should keep their photon pairs for the next round. That is, if Alice and
Bob get the phase shift $\theta_{1}$, they obtain the state of the two photon
pairs $(|V\rangle_{c1}|H\rangle_{d1}|V\rangle_{c2}|H\rangle_{d2}
+|H\rangle_{c1}|V\rangle_{d1}|H\rangle_{c2}|V\rangle_{d2})$. If they
both get the phase shift $\theta_{0}$ (condition $\theta_{2}=\theta_{0}+2\pi$), they get the state
$(|V\rangle_{c1}|H\rangle_{d1}|H\rangle_{c2}|V\rangle_{d2}
+|H\rangle_{c1}|V\rangle_{d1}|V\rangle_{c2}|H\rangle_{d2})$.
Therefore, they make an operation
$\hat{\sigma}_{x}=|V\rangle\langle H|+|H\rangle\langle V|$ on
$\hat{c}_{2}$ and $\hat{d}_{2}$ to obtain the state
$(|V\rangle_{c1}|H\rangle_{d1}|V\rangle_{c2}|H\rangle_{d2}
+|H\rangle_{c1}|V\rangle_{d1}|H\rangle_{c2}|V\rangle_{d2})$.
Subsequently, Alice and Bob make a measurement with the diagonal
basis $\{|\pm\rangle=\frac{1}{\sqrt{2}}(|H\rangle\pm|V\rangle)\}$ on
$\hat{c}_{2}$ and $\hat{d}_{2}$. If they both obtain the results
$|+\rangle$ or $|-\rangle$, the state of microwave-photon pairs
$\hat{c}_{1}\hat{d}_{1}$ becomes $|\Psi^{\dag}\rangle_{c1d1}$.
Otherwise, they should make the operation
$\hat{\sigma}_{z}=|H\rangle\langle H|-|V\rangle\langle V|$
on the microwave photon $\hat{c}_{1}$ to obtain the state
$|\Psi^{\dag}\rangle_{c_1d_1}$.

After the operations, Alice and Bob can obtain their nonlocal
entangled state of microwave-photon pairs with more purity.
In the ideal model, the fidelity of the
remaining microwave-photon pairs is given by
\begin{eqnarray}        \label{eq14}
f_{ideal}=\frac{f^{2}}{f^{2}+(1-f)^{2}}.
\end{eqnarray}

\begin{figure*}[!ht]%[tpb]
\begin{center}
\includegraphics[width=16cm,angle=0]{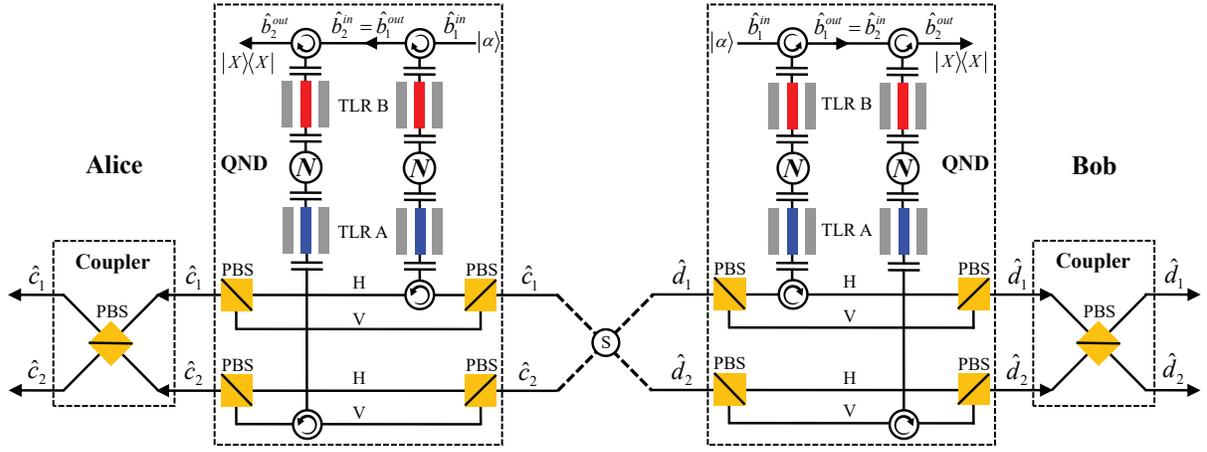}
\caption{ Schematic diagram of our EPP for polarization-spatial
entangled microwave-photon pairs. $S$ is the entanglement source for
generating polarization-spatial entangled microwave-photon pairs.
The two big dashed boxes are two  polarization parity-check QND detectors. The two
small dashed boxes are two couplers with the same PBS for microwave
photons.} \label{figeppps}
\end{center}
\end{figure*}

\begin{table}
  \begin{center}
\caption{ Corresponding relation between the states of the signal
light in storage resonators and the phase shifts by choosing two same cross-Kerr systems
in each QND detector.}
    \begin{tabular}{lccccccccccccccccccccccccccccc}
      \hline\hline
State $|\hat{a}_{1}\rangle|\hat{a}_{2}\rangle$&&&&&&&&&&&&&&&&&&&&&&&&&&&&& Total phase shift \\ \hline
$|0\rangle|0\rangle$ &&&&&&&&&&&&&&&&&&&&&&&&&&&&& $\theta_{0}$ \\
$|1\rangle|0\rangle/|0\rangle|1\rangle$ &&&&&&&&&&&&&&&&&&&&&&&&&&&&& $\theta_{1}$ \\
$|1\rangle|1\rangle$ &&&&&&&&&&&&&&&&&&&&&&&&&&&&& $\theta_{2}$ \\
$|2\rangle|0\rangle/|0\rangle|2\rangle$ &&&&&&&&&&&&&&&&&&&&&&&&&&&&& $\theta_{3}$
\\ \hline\hline
    \end{tabular}\label{Tab3}
  \end{center}
\end{table}

\subsection{Entanglement purification for polarization-spatial entangled microwave-photon pairs}
\label{sec32}

The polarization-spatial entangled states are widely used in quantum
communication as they can be produced by parametric down-conversion
naturally in experiment. Therefore, considering the situation for
polarization-spatial entangled microwave-photon pairs is very
necessary. For a pair polarization-spatial entangled microwave
photons, the state is given by
$(\hat{c}_{1H}^{\dag}\hat{d}_{1H}^{\dag}+\hat{c}_{1V}^{\dag}\hat{d}_{1V}^{\dag}
+\hat{c}_{2H}^{\dag}\hat{d}_{2H}^{\dag}+\hat{c}_{2V}^{\dag}\hat{d}_{2V}^{\dag})|0\rangle$.
Therefore, for the four-photon state, it can be described by
$(\hat{c}_{1H}^{\dag}\hat{d}_{1H}^{\dag}+\hat{c}_{1V}^{\dag}\hat{d}_{1V}^{\dag}
+\hat{c}_{2H}^{\dag}\hat{d}_{2H}^{\dag}+\hat{c}_{2V}^{\dag}\hat{d}_{2V}^{\dag})^{2}|0\rangle$.
The detailed schematic diagram of our EPP for those two situations
is shown in Fig.~\ref{figeppps}. Here the states $|H\rangle$ and
$|V\rangle$ are translated to $|1\rangle$ and $|0\rangle$ in
$\hat{c}_{1}\hat{d}_{1}$ mode, respectively. In
$\hat{c}_{2}\hat{d}_{2}$ mode, the corresponding relations are
opposite. Here, we choose the two same cross-Kerr systems in each
QND detector and the corresponding phase shifts are given in
Table~\ref{Tab3}. We assume all the four angles are different in
this section.

First, we consider the case where there is a pair of
polarization-spatial entangled microwave photons. This time, it
is just an ideal microwave-photon pair. After it passes through the
QND detectors, the state composed of the microwave-photon pair and
the probe light is given by
\begin{eqnarray}        \label{eq15}
\Longrightarrow\!\!\!\!&&(\hat{c}_{1H}^{\dag}\hat{d}_{1H}^{\dag}
+\hat{c}_{2V}^{\dag}\hat{d}_{2V}^{\dag})|0\rangle|\alpha
e^{i\theta_{1}}\rangle_{c}
|\alpha e^{i\theta_{1}}\rangle_{d}\nonumber\\
&&+(\hat{c}_{1V}^{\dag}\hat{d}_{1V}^{\dag}+\hat{c}_{2H}^{\dag}\hat{d}_{2H}^{\dag})
|0\rangle|\alpha e^{i\theta_{0}}\rangle_{c}|\alpha
e^{i\theta_{0}}\rangle_{d}.\;\;\;\;\;\;
\end{eqnarray}
When Alice and Bob get the same phase shift $\theta_{1}$ via an \emph{X}
homodyne measurement on their probe lights, they obtain the state of
their microwave-photon pair
$(\hat{c}_{1H}^{\dag}\hat{d}_{1H}^{\dag}+\hat{c}_{2V}^{\dag}\hat{d}_{2V}^{\dag})|0\rangle$.
After passing through the couplers, their photon pair will appear at
the modes $\hat{c}_{2}\hat{d}_{2}$. When Alice and Bob get the same
phase shift $\theta_{0}$, they obtain the state
$(\hat{c}_{1V}^{\dag}\hat{d}_{1V}^{\dag}+\hat{c}_{2H}^{\dag}\hat{d}_{2H}^{\dag})|0\rangle$,
and then their photon pair will appear at the output modes
$\hat{c}_{1}\hat{d}_{1}$. When the bit-flipping error occurs, the
state becomes
$(\hat{c}_{1V}^{\dag}\hat{d}_{1H}^{\dag}+\hat{c}_{2H}^{\dag}\hat{d}_{2V}^{\dag}
+\hat{c}_{1H}^{\dag}\hat{d}_{1V}^{\dag}+\hat{c}_{2V}^{\dag}\hat{d}_{2H}^{\dag})|0\rangle$.
With the QND measurement, the state of the photon pair evolves to
\begin{eqnarray}        \label{eq16}
\Longrightarrow\!\!\!\!&&(\hat{c}_{1V}^{\dag}\hat{d}_{1H}^{\dag}
+\hat{c}_{2H}^{\dag}\hat{d}_{2V}^{\dag})|0\rangle|\alpha
e^{i\theta_{0}}\rangle_{c}
|\alpha e^{i\theta_{1}}\rangle_{d}\nonumber\\
&&+(\hat{c}_{1H}^{\dag}\hat{d}_{1V}^{\dag}+\hat{c}_{2V}^{\dag}\hat{d}_{2H}^{\dag})
|0\rangle|\alpha e^{i\theta_{1}}\rangle_{c}|\alpha
e^{i\theta_{0}}\rangle_{d}.\;\;\;\;\;\;
\end{eqnarray}
Alice and Bob will get the different results $\theta_{0}$ and
$\theta_{1}$. They should perform a bit-flip operation of
polarization $\hat{\sigma}_{x}=|V\rangle\langle
H|+|H\rangle\langle V|$ on photon $\hat{c}_1$ to obtain the state
$(\hat{c}_{H}^{\dag}\hat{d}_{H}^{\dag}+\hat{c}_{V}^{\dag}\hat{d}_{V}^{\dag})|0\rangle$.

Second, we consider the case where there are two pairs of
polarization-spatial entangled microwave photons. With no
decoherence, the state of the two photon pairs is expressed as
$(\hat{c}_{1H}^{\dag}\hat{d}_{1H}^{\dag}+\hat{c}_{1V}^{\dag}\hat{d}_{1V}^{\dag}
+\hat{c}_{2H}^{\dag}\hat{d}_{2H}^{\dag}+\hat{c}_{2V}^{\dag}\hat{d}_{2V}^{\dag})^{2}|0\rangle$.
After the QND measurements are performed by Alice and Bob, the state
of the whole system composed of the photon pair and the probe lights
is
\begin{eqnarray}        \label{eq17}
\Longrightarrow\!\!\!\!&&[(\hat{c}_{1H}^{\dag}\hat{d}_{1H}^{\dag})^{2}
+(\hat{c}_{2V}^{\dag}\hat{d}_{2V}^{\dag})^{2}]|0\rangle
|\alpha e^{i\theta_{3}}\rangle_{c}|\alpha e^{i\theta_{3}}\rangle_{d}\nonumber\\
&&+2\hat{c}_{1H}^{\dag}\hat{d}_{1H}^{\dag}\hat{c}_{2V}^{\dag}\hat{d}_{2V}^{\dag}|0\rangle
|\alpha e^{i\theta_{2}}\rangle_{c}|\alpha e^{i\theta_{2}}\rangle_{d}\nonumber\\
&&+(\hat{c}_{1V}^{\dag}\hat{d}_{1V}^{\dag}+\hat{c}_{2H}^{\dag}\hat{d}_{2H}^{\dag})^{2}
|0\rangle|\alpha e^{i\theta_{0}}\rangle_{c}|\alpha e^{i\theta_{0}}\rangle_{d}\nonumber\\
&&+2(\hat{c}_{1H}^{\dag}\hat{d}_{1H}^{\dag}+\hat{c}_{2V}^{\dag}\hat{d}_{2V}^{\dag})
(\hat{c}_{1V}^{\dag}\hat{d}_{1V}^{\dag}+\hat{c}_{2H}^{\dag}\hat{d}_{2H}^{\dag})\;\;\;\;\;\;\;\nonumber\\
&&\otimes|0\rangle|\alpha e^{i\theta_{1}}\rangle_{c}|\alpha
e^{i\theta_{1}}\rangle_{d}.
\end{eqnarray}
Alice and Bob will get four results with $\theta_{0}$,
$\theta_{1}$, $\theta_{2}$, and $\theta_{3}$ which  correspond to the states
$(\hat{c}_{1V}^{\dag}\hat{d}_{1V}^{\dag}+\hat{c}_{2H}^{\dag}\hat{d}_{2H}^{\dag})^{2}|0\rangle$,
$(\hat{c}_{1H}^{\dag}\hat{d}_{1H}^{\dag}+\hat{c}_{2V}^{\dag}\hat{d}_{2V}^{\dag})
(\hat{c}_{1V}^{\dag}\hat{d}_{1V}^{\dag}+\hat{c}_{2H}^{\dag}\hat{d}_{2H}^{\dag})|0\rangle$,
$\hat{c}_{1H}^{\dag}\hat{d}_{1H}^{\dag}\hat{c}_{2V}^{\dag}\hat{d}_{2V}^{\dag}|0\rangle$,
and $(\hat{c}_{1H}^{\dag}\hat{d}_{1H}^{\dag})^{2}+(\hat{c}_{2V}^{\dag}\hat{d}_{2V}^{\dag})^{2}|0\rangle$,
respectively. After the photons pass through the couplers, the states
$[(\hat{c}_{1H}^{\dag}\hat{d}_{1H}^{\dag})^{2}+(\hat{c}_{2V}^{\dag}\hat{d}_{2V}^{\dag})^{2}]|0\rangle$
and
$(\hat{c}_{1V}^{\dag}\hat{d}_{1V}^{\dag}+\hat{c}_{2H}^{\dag}\hat{d}_{2H}^{\dag})^{2}|0\rangle$
will appear at $\hat{c}_{2}\hat{d}_{2}$ and $\hat{c}_{1}\hat{d}_{1}$, respectively.
The two photon pairs
$(\hat{c}_{1H}^{\dag}\hat{d}_{1H}^{\dag}+\hat{c}_{2V}^{\dag}\hat{d}_{2V}^{\dag})
(\hat{c}_{1V}^{\dag}\hat{d}_{1V}^{\dag}+\hat{c}_{2H}^{\dag}\hat{d}_{2H}^{\dag})|0\rangle$
will be divided into $\hat{c}_{1}\hat{d}_{1}$ and
$\hat{c}_{2}\hat{d}_{2}$, respectively.

When the bit-flipping error occurs, there will be two situations.
The first situation is that only one of two microwave-photon pairs
has an error, and the state of the two photon pairs becomes
$(\hat{c}_{1H}^{\dag}\hat{d}_{1H}^{\dag}+\hat{c}_{1V}^{\dag}\hat{d}_{1V}^{\dag}
+\hat{c}_{2H}^{\dag}\hat{d}_{2H}^{\dag}+\hat{c}_{2V}^{\dag}\hat{d}_{2V}^{\dag})
(\hat{c}_{1V}^{\dag}\hat{d}_{1H}^{\dag}+\hat{c}_{1H}^{\dag}\hat{d}_{1V}^{\dag}
+\hat{c}_{2V}^{\dag}\hat{d}_{2H}^{\dag}+\hat{c}_{2H}^{\dag}\hat{d}_{2V}^{\dag})|0\rangle$.
Therefore, with the QND detector, the composite system composed of
the two photon pairs and the two probe lights evolves to
\begin{eqnarray}        \label{eqoneerror}
\Longrightarrow&&(\hat{c}_{1H}^{\dag}\hat{d}_{1H}^{\dag}\hat{c}_{1V}^{\dag}\hat{d}_{1H}^{\dag}
+\hat{c}_{2V}^{\dag}\hat{d}_{2V}^{\dag}\hat{c}_{2H}^{\dag}\hat{d}_{2V}^{\dag})\nonumber\\
&&\otimes|0\rangle|\alpha e^{i\theta_{1}}\rangle_{c}|\alpha e^{i\theta_{3}}\rangle_{d}\nonumber\\
&&+(\hat{c}_{1H}^{\dag}\hat{d}_{1H}^{\dag}\hat{c}_{2H}^{\dag}\hat{d}_{2V}^{\dag}
+\hat{c}_{2V}^{\dag}\hat{d}_{2V}^{\dag}\hat{c}_{1V}^{\dag}\hat{d}_{1H}^{\dag})\nonumber\\
&&\otimes|0\rangle|\alpha e^{i\theta_{1}}\rangle_{c}|\alpha e^{i\theta_{2}}\rangle_{d}\nonumber\\
&&+(\hat{c}_{1H}^{\dag}\hat{d}_{1H}^{\dag}\hat{c}_{1H}^{\dag}\hat{d}_{1V}^{\dag}
+\hat{c}_{2V}^{\dag}\hat{d}_{2V}^{\dag}\hat{c}_{2V}^{\dag}\hat{d}_{2H}^{\dag})\nonumber\\
&&\otimes|0\rangle|\alpha e^{i\theta_{3}}\rangle_{c}|\alpha e^{i\theta_{1}}\rangle_{d}\nonumber\\
&&+(\hat{c}_{1H}^{\dag}\hat{d}_{1H}^{\dag}\hat{c}_{2V}^{\dag}\hat{d}_{2H}^{\dag}
+\hat{c}_{2V}^{\dag}\hat{d}_{2V}^{\dag}\hat{c}_{1H}^{\dag}\hat{d}_{1V}^{\dag})\nonumber\\
&&\otimes|0\rangle|\alpha e^{i\theta_{2}}\rangle_{c}|\alpha e^{i\theta_{1}}\rangle_{d}\nonumber\\
&&+(\hat{c}_{1V}^{\dag}\hat{d}_{1V}^{\dag}+\hat{c}_{2H}^{\dag}\hat{d}_{2H}^{\dag})
(\hat{c}_{1V}^{\dag}\hat{d}_{1H}^{\dag}+\hat{c}_{2H}^{\dag}\hat{d}_{2V}^{\dag})\nonumber\\
&&\otimes|0\rangle|\alpha e^{i\theta_{0}}\rangle_{c}|\alpha e^{i\theta_{1}}\rangle_{d}\nonumber\\
&&+(\hat{c}_{1V}^{\dag}\hat{d}_{1V}^{\dag}+\hat{c}_{2H}^{\dag}\hat{d}_{2H}^{\dag})
(\hat{c}_{1H}^{\dag}\hat{d}_{1V}^{\dag}+\hat{c}_{2V}^{\dag}\hat{d}_{2H}^{\dag})\nonumber\\
&&\otimes|0\rangle|\alpha e^{i\theta_{1}}\rangle_{c}|\alpha e^{i\theta_{0}}\rangle_{d}
\end{eqnarray}
Analyzing from the result in Eq.~(\ref{eqoneerror}), Alice and Bob
know there exists an error in one pair when they get the different
phase shifts. Alice and Bob should discard this result because they
cannot get the state
$(\hat{c}_{1H}^{\dag}\hat{d}_{1H}^{\dag}+\hat{c}_{2V}^{\dag}\hat{d}_{2V}^{\dag})|0\rangle$
at $\hat{c}_{1}\hat{d}_{1}$ and $\hat{c}_{2}\hat{d}_{2}$.

The second one is the bit-flipping error taking place on both the
two microwave photon pairs. The state becomes
$(\hat{c}_{1V}^{\dag}\hat{d}_{1H}^{\dag}+\hat{c}_{1H}^{\dag}\hat{d}_{1V}^{\dag}
+\hat{c}_{2V}^{\dag}\hat{d}_{2H}^{\dag}+\hat{c}_{2H}^{\dag}\hat{d}_{2V}^{\dag})^{2}|0\rangle$.
When the two pairs pass the QND detectors, the state of the system will evolve to
\begin{eqnarray}   \label{eq19}
\Longrightarrow\!\!\!\!&&[(\hat{c}_{1H}^{\dag}\hat{d}_{1V}^{\dag})^{2}
+(\hat{c}_{2V}^{\dag}\hat{d}_{2H}^{\dag})^{2}]|0\rangle
|\alpha e^{i\theta_{3}}\rangle_{c}|\alpha e^{i\theta_{0}}\rangle_{d}\nonumber\\
&&+2\hat{c}_{1H}^{\dag}\hat{d}_{1V}^{\dag}\hat{c}_{2V}^{\dag}\hat{d}_{2H}^{\dag}|0\rangle
|\alpha e^{i\theta_{2}}\rangle_{c}|\alpha e^{i\theta_{0}}\rangle_{d}\nonumber\\
&&+[(\hat{c}_{1V}^{\dag}\hat{d}_{1H}^{\dag})^{2}+(\hat{c}_{2H}^{\dag}\hat{d}_{2V}^{\dag})^{2}]
|0\rangle|\alpha e^{i\theta_{0}}\rangle_{c}|\alpha e^{i\theta_{3}}\rangle_{d}\nonumber\\
&&+2\hat{c}_{1V}^{\dag}\hat{d}_{1H}^{\dag}\hat{c}_{2H}^{\dag}\hat{d}_{2V}^{\dag}
|0\rangle|\alpha e^{i\theta_{0}}\rangle_{c}|\alpha e^{i\theta_{2}}\rangle_{d}\nonumber\\
&&+2(\hat{c}_{1V}^{\dag}\hat{d}_{1H}^{\dag}+\hat{c}_{2H}^{\dag}\hat{d}_{2V}^{\dag})
(\hat{c}_{1H}^{\dag}\hat{d}_{1V}^{\dag}+\hat{c}_{2V}^{\dag}\hat{d}_{2H}^{\dag})\;\;\;\;\;\;\;\;\nonumber\\
&&\otimes|0\rangle|\alpha e^{i\theta_{1}}\rangle_{c}|\alpha
e^{i\theta_{1}}\rangle_{d}.
\end{eqnarray}
Alice and Bob get the five results in which four results have
different phase shifts and the other has the same phase shift. For
the different phase shifts, they should discard the photon pairs because
the pairs will appear in the same spatial mode. When Alice and Bob
get the same phase shift, they cannot distinguish it from the situation
with no error and they should keep the pairs. Then Alice and Bob can
continue to purify the states by using the protocol presented for
ideal entanglement sources discussed in Sec.~\ref{sec31}.

\subsection{Parameters and fidelity for quantum nondemolition detector}
\label{sec33}

The cross-Kerr effect is induced by coupling two TLRs to a
superconducting molecule as shown in Fig.~\ref{fig1}. According to
the previous works \cite{HZhangPRA2017,YHu}, we choose the
parameters of this superconducting system as  $E_{c}/2\pi=0.5$ GHz,
$E_{J}/2\pi=16$ GHz, and $E_{m}/2\pi=0.2$ GHz. The two coupling
strengthes between the molecule and the TLRs are equal with
$\text{g}_{1}/2\pi\sim\text{g}_{2}/2\pi\sim 300$ MHz. The classical
pump field strength $\Omega_{c}$ and the detuning $\delta_{2}$ are
designed to $\Omega_{c}/2\pi\sim\delta_{2}/2\pi\sim 1.5$ GHz.
Therefore, the cross-Kerr effect coefficient in our scheme is
$|\chi|/2\pi\sim 2.4$ MHz. A recent experiment \cite{ETHolland}
demonstrated a state-dependent shift $|\chi_{sc}|/2\pi=2.59\pm 0.06$
MHz between two cavities in circuit QED.

According to the parameters chosen above, we calculate the angle of
the coherent state in Table~\ref{Tab3}. The angles of states
$|00\rangle$, $|10\rangle/|01\rangle$, $|11\rangle$ and
$|20\rangle/|02\rangle$ are $\theta_{0}=0$, $\theta_{1}\approx 0.6$,
$\theta_{2}\approx 1.2$ and $\theta_{3}\approx 1.1$, respectively.
Here, we choose the decay rate with $\kappa_{2}^{-1}\approx 10$ ns in our calculation.
If we require that the minimal error probability is less than 0.01,
according to the Eq.~(\ref{erfc}), then $X_{d}$ should satisfy
$X_{d}>4.5$. Therefore, when the $\alpha$ satisfies $\alpha>24.7$,
all the angles can be distinguished.  Here, the cross-Kerr effect is
weak, the phase shifts in Table~\ref{Tab2} can not satisfy the
condition $\theta_{2}=\theta_{0}+2\pi$. Therefore, in this physical
system, Alice and Bob should make operations under the situation
with different phase shifts.

\begin{figure}[!ht]%[tpb]
\begin{center}
\includegraphics[width=8.0cm,angle=0]{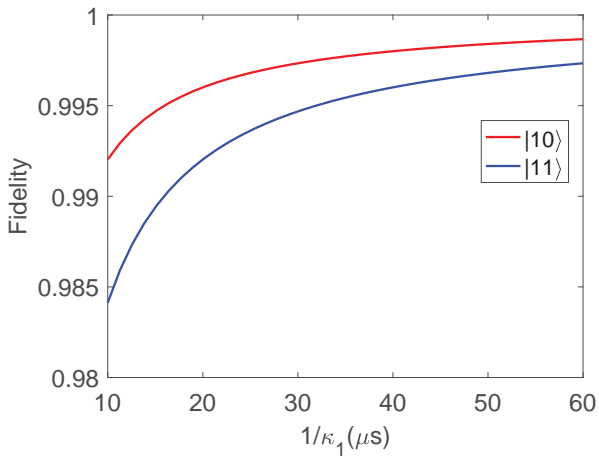}
\caption{ The fidelity of the state in the storage resonators with dissipation for different decay rates
$\kappa_{1}^{-1}$. Here the decay rate of the readout resonators is
$\kappa_{2}^{-1}\sim 10$ ns.} \label{figka1}
\end{center}
\end{figure}

\begin{figure}[!ht]%[tpb]
\begin{center}
\includegraphics[width=8.0cm,angle=0]{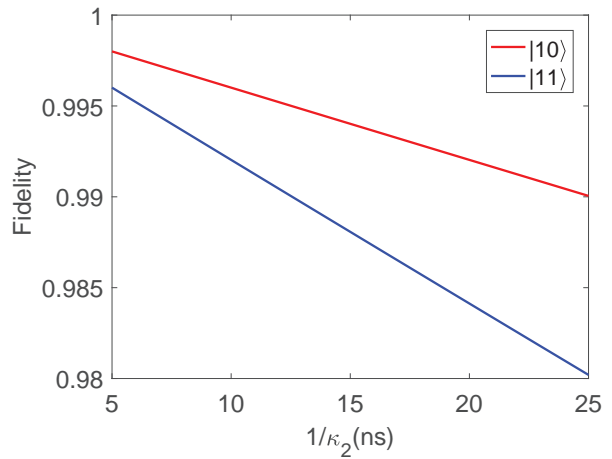}
\caption{ The fidelity of the state in the storage resonators with dissipation for
different decay rates of the readout resonators $\kappa_{2}^{-1}$. Here the decay rate of the
storage resonators is $\kappa_{1}^{-1}\sim 20$ $\mu$s.} \label{figka2}
\end{center}
\end{figure}

In practice, the number of microwave photons will decrease due to
the dissipation of storage resonators. The dynamics of the quantum system with
dissipation is described by the master equation in Lindblad form
given by
\begin{eqnarray}        \label{eq20}
\frac{d\hat{\rho}(t)}{dt}\!=\!i\left[\hat{\rho}(t),\hat{H}\right]\!
+\!\kappa_{1}\hat{L}\left[\hat{a_{1}}\right]\hat{\rho}(t)
\!+\!\kappa_{1}\hat{L}\left[\hat{a_{2}}\right]\hat{\rho}(t),\;\;
\end{eqnarray}
where $\hat{\rho}(t)$ and $\hat{H}$ are the density matrix and
the Hamiltonian of the system, respectively. The symbols
$\kappa_{1}$ represents the decay rates of TLR
\emph{A}. The superoperator $\hat{L}$ with the rule
$\hat{L}[\hat{o}]\hat{\rho}=(2\hat{o}\hat{\rho}\hat{o}^{\dag}
-\hat{o}^{\dag}\hat{o}\hat{\rho}-\hat{\rho}\hat{o}^{\dag}\hat{o})/2$
represents the influence of the dissipation.
Here, we assume that the QND measurement is ideal and that the probe light has no influence on the states in storage resonators. The fidelity is influenced by the leakage of
the resonator. The formula of fidelity is $F=\langle\psi_{id}|\hat{\rho}(t)|\psi_{id}\rangle$,
where the ideal state $|\psi_{id}\rangle$ is the initial state here. We consider the two initial states with $|10\rangle$ and $|11\rangle$. State $|nm\rangle$ represents $n$ and $m$ photons in storage resonator $A_{1}$ and $A_{2}$, respectively.
We calculate the fidelities of the states in storage resonators at the end of the measuring time. In the rotating frame, the Hamiltonian of the resonator is zero. We choose the total measuring time of the cascade system with $\tau\sim 8/\kappa_{2}$.
The decay rate of the readout resonator keeps
$\kappa_{2}^{-1}\sim 10$ ns in the whole process. The fidelities are
proportional to $\kappa_{1}^{-1}$ in Fig.~\ref{figka1}, which
indicates that the large storage time (the better resonator) can
protect the microwave photons from dissipation. Then we plot the influences from
the different $\kappa_{2}$ in Fig.~\ref{figka2}. The parameter here is $\kappa_{1}^{-1}\sim20 \mu$s.
Contrary to Fig.~\ref{figka1}, the fidelities are inversely proportional to
$\kappa_{2}^{-1}$ in Fig.~\ref{figka2}. The large $\kappa_{2}^{-1}$
means a long measuring time. Therefore, as  $\kappa_{2}^{-1}$
becomes large, it will result in more total dissipation and the
fidelity becomes lower.

\section{summary}
\label{sec4}

In summary, we have proposed a physically feasible polarization EPP
for the entangled state of nonlocal microwave photons in circuit
QED. Our EPP includes two processes. The first process is used to purify
the polarization entanglement state generated by the ideal
entanglement sources and the second process is used for
polarization-spatial entangled microwave-photon pairs. In our EPP,
we design the polarization parity-check QND detectors to realize the
postselection of microwave-photon quantum states. According to the
phase shifts of the probe light hold by the two remote parties in
quantum communication, say Alice and Bob, the parties can
distinguish whether the error takes place and then correct it. We
implement the QND measurement based on the cross-Kerr effect induced by
coupling the two TLRs to a superconducting molecule. Our work can
improve the practical application of microwave-based quantum
communication. For example, quantum repeaters are the indispensable
parts in long-distance quantum communication. Due to the unavoidable
influence of the environment in the processes of transmission and
storage, the nonlocal near maximally entangled state generated
between every two neighboring nodes and used as the quantum channel
in a quantum repeater may turn into a mixed entangled state.
Therefore, our purification protocol can be used here. Also, in the
actual situation of satellite quantum communication, when the
microwave signals pass through the aerosphere from the quantum
satellite to the ground, the pure maximally entangled
microwave-photon state may become the mixed one due to the influence
of environment in the process of satellite quantum communication. To keep the communication efficient, the parties can use our
EPP to purify the mixed entangled microwave-photon state to improve
its fidelity.

\section*{ACKNOWLEDGMENTS}

We thank Guan-Yu Wang, Jing Qiu, and Zhi-Sheng Yang for helpful
discussions. This work is supported by the National Natural Science
Foundation of China under Grants No. 11674033, No. 11474026, and No.
11474027,  the Fundamental Research Funds for the Central
Universities under Grant No. 2015KJJCA01, and the National Key
Basic Research Program of China under Grant No. 2013CB922000.\\

%\end{CJK*}


\begin{thebibliography}{1}

\bibitem{CHBennet1993} C. H. Bennett, G. Brassard, C. Crepeau, R. Jozsa, A. Peres,
and W. K. Wootters, Teleporting an Unknown Quantum State via Dual Classical
and Einstein-Podolsky-Rosen Channels, Phys. Rev. Lett. \textbf{70}, 1895 (1993).

\bibitem{CHBennet1992} C. H. Bennett and S. J. Wiesner, Communication via
One- and Two-Particle Operators on Einstein-Podolsky-Rosen states,
Phys. Rev. Lett. \textbf{69}, 2881 (1992).

\bibitem{superdense} X. S. Liu, G. L. Long, D. M. Tong, and F. Li,
General scheme for superdense coding between multiparties,
Phys. Rev. A \textbf{65}, 022304 (2002).

\bibitem{AKEkert} A. K. Ekert, Quantum Cryptography Based on Bell's Theorem,
Phys. Rev. Lett. \textbf{67}, 661 (1991).

\bibitem{bbm92} C. H. Bennett, G. Brassard, and N. D. Mermin,
Quantum Cryptography Without Bell's Theorem, Phys. Rev. Lett. \textbf{68}, 557 (1992).


\bibitem{lixhqkdpra} X. H. Li, F. G. Deng, and H. Y. Zhou,
Efficient quantum key distribution over a collective noise channel,
Phys. Rev. A \textbf{78}, 022321 (2008).


\bibitem{MHillery} M. Hillery, V. Bu\v{z}ek, and A. Berthiaume,
Quantum secret sharing, Phys. Rev. A \textbf{59}, 1829 (1999).





\bibitem{longliupra} G. L. Long and X. S. Liu,
Theoretically efficient high-capacity quantum-key-distribution scheme,
Phys. Rev. A \textbf{65}, 032302 (2002).

\bibitem{FGDeng2003} F. G. Deng, G. L. Long, and X. S. Liu,
Two-step quantum direct communication protocol using the Einstein-Podolsky-Rosen pair block,
Phys. Rev. A \textbf{68}, 042317 (2003).


\bibitem{QSDCWangC} C. Wang, F. G. Deng, Y. S. Li,  X. S. Liu,  and G. L.
Long, Quantum secure direct communication with high-dimension
quantum superdense coding, Phys. Rev. A  \textbf{71}, 044305 (2005).


\bibitem{twostepexp} W. Zhang, D. S. Ding, Y. B. Sheng, L. Zhou, B. S. Shi, and G. C Guo,
Quantum secure direct communication with quantum memory,
Phys. Rev. Lett. \textbf{118}, 220501 (2017).




%\bibitem{faithful} X. H. Li, F. G. Deng, and H. Y. Zhou,
%Faithful qubit transmission against collective noise without ancillary qubits,
%Appl. Phys. Lett. \textbf{91}, 144101 (2007).



\bibitem{DFS1}  Z. D. Walton, A. F. Abouraddy, A. V. Sergienko,
B. E. A. Saleh, and M. C. Teich, Decoherence-Free Fubspaces in Quantum Key Distribution,
Phys. Rev. Lett. \textbf{91}, 087901 (2003).


\bibitem{DFS2} J. C. Boileau, D. Gottesman, R. Laflamme, D. Poulin,
and R. W. Spekkens, Robust Polarization-Based Quantum Key Distribution
Over a Collective-Noise Channel, Phys. Rev. Lett. \textbf{92}, 017901 (2004).


\bibitem{DFS3} J. C. Boileau, R. Laflamme, M. Laforest, and C. R. Myers,
Robust Quantum Communication Using a Polarization-Entangled Photon Pair,
Phys. Rev. Lett. \textbf{93}, 220501 (2004).








%%%%%%%%%%%%%%%%%%%  ECP

\bibitem{CHBennett1996} C. H. Bennett, H. J. Bernstein, S. Popescu, and B. Schumacher,
Concentrating partial entanglement by local operations, Phys. Rev. A \textbf{53}, 2046 (1996).

%\bibitem{ZZhao014301} Z. Zhao, J. W. Pan, and M. S. Zhan,
%Practical scheme for entanglement concentration, Phys. Rev. A \textbf{64}, 014301 (2001).
%
%\bibitem{TYamamoto012304} T. Yamamoto, M. Koashi, and N. Imoto,
%Concentration and purification scheme for two partially entangled photon pairs,
%Phys. Rev. A \textbf{64}, 012304 (2001).



\bibitem{YBSheng2008} Y. B. Sheng, F. G. Deng, and H. Y. Zhou,
Nonlocal entanglement concentration scheme for partially entangled
multipartite systems with nonlinear optics, Phys. Rev. A
\textbf{77}, 062325 (2008).




%\bibitem{BCRenLPLECP}B. C. Ren and F. G. Deng,
%Hyperentanglement purification and concentration assisted by
%diamond NV centers inside photonic crystal cavities, Laser Phys. Lett. \textbf{10}, 115201 (2013).
%
%\bibitem{TLiatom} T.  Li,  G. J. Yang,  and F. G. Deng,
%Entanglement distillation for quantum communication network with atomic-ensemble memories,
%Opt. Express \textbf{22}, 23897 (2014).


%
%
%\bibitem{SBose194} S. Bose, V. Vedral, P. L. Knight,
%Purification via entanglement swapping and conserved entanglement,
%Phys. Rev. A \textbf{60}, 194 (1999).
%
%\bibitem{YBSheng012307} Y. B. Sheng, L. Zhou, S. M. Zhao, and B. Y. Zheng,
%Efficient single-photon-assisted entanglement concentration for partially
%entangled photon pairs, Phys. Rev. A \textbf{85}, 012307 (2012).
%
%\bibitem{FGDeng022311} F. G. Deng, Optimal nonlocal multipartite
%entanglement concentration based on projection measurements,
%Phys. Rev. A \textbf{85}, 022311 (2012).

\bibitem{BCRen012302} B. C. Ren, F. F. Du, and F. G. Deng,
Hyperentanglement concentration for two-photon four-qubit systems with
linear optics, Phys. Rev. A \textbf{88}, 012302 (2013).


%
%\bibitem{BCRen6547} B. C. Ren and G. L. Long, General hyperentanglement
%concentration for photon systems assisted by
%    quantum-dot spins inside optical microcavities, Opt. Express \textbf{22}, 6547 (2014).%-6561


%\bibitem{BCRen16444} B. C. Ren and G. L. Long,
%Highly efficient hyperentanglement concentration with two steps
%assisted by quantum swap gates, Sci. Rep. \textbf{5}, 16444  (2015).


%\bibitem{XHLi125201} X. H. Li and S. Ghose, Hyperconcentration
%for multipartite entanglement via linear optics, Laser phys. Lett. \textbf{11}, 125201 (2014).



%\bibitem{XHLi3550} X. H. Li and S. Ghose,
%Efficient hyperconcentration of nonlocal multipartite entanglement via the
%    cross-Kerr nonlinearity, Opt. Express \textbf{23}, 3550 (2015).%-3562


\bibitem{Lixhpraecp} X. H. Li and S. Ghose,
Hyperentanglement concentration for time-bin and polarization hyperentangled photons,
Phys. Rev. A \textbf{91}, 062302 (2015).






%
%\bibitem{WangECPOE15} T. J. Wang, L. L. Liu, R. Zhang, C. Cao, and C. Wang,
%One-step hyperentanglement purification and hyperdistillation with linear optics,
%Opt. Express \textbf{23}, 009284 (2015).
%
%\bibitem{WangECPAP16} C. Cao, T. J. Wang, S. C. Mi, R. Zhang, C. Wang,
%Nonlocal hyperconcentration on entangled photons
%    using photonic module system, Ann. Phys. \textbf{369}, 128 (2016).%--138
%

%
%\bibitem{CWangECPPRA12} C. Wang,
%Efficient entanglement concentration for partially entangled electrons using
%a quantum-dot and microcavity coupled system, Phys. Rev. A \textbf{86}, 012323 (2012).



\bibitem{HZhangPRA2017} H. Zhang and H. Wang,
Entanglement concentration of microwave photons based on the Kerr effect in circuit QED,
Phys. Rev. A \textbf{95}, 052314 (2017).








%%%%%%%%%%%%%%%%%%%%%%%%%%  EPP


\bibitem{EPP1} C. H. Bennett, G. Brassard, S. Popescu, B. Schumacher, J. A. Smolin,
and W. K. Wootters,
 Purification of Noise Entanglement and Faithful Teleportation via Noisy Channels,
 Phys. Rev. Lett. \textbf{76}, 722 (1996).%--725


\bibitem{DDeutschPRL1996} D. Deutsch, A. Ekert, R. Jozsa, C. Macchiavello,
S. Popescu, and A. Sanpera, Quantum Privacy Amplification and
the Security of Quantum Cryptography over Noisy Channels,
Phys. Rev. Lett. \textbf{77}, 2818 (1996).


%\bibitem{MMuraoPRA1998}M. Murao, M. B. Plenio, S. Popescu, V. Vedral,
%and P. L. Knight, Multiparticle entanglement purification protocols,
%Phys. Rev. A \textbf{57}, R4075(R) (1998).
%
%\bibitem{LMDuanPRL2000} L. M. Duan, G. Giedke, J. I. Cirac, and P. Zoller,
%Entanglement Purification of Gaussian Continuous Variable Quantum States,
%Phys. Rev. Lett. \textbf{84}, 4002 (2000).




\bibitem{JWPannature2001} J. W. Pan, C. Simon, C. Brukner, and A. Zelinger,
Entanglement purification for quantum communication, Nature (London) \textbf{410}, 1067 (2001).


\bibitem{Simon2002} C. Simon and J.W. Pan,
Polarization Entanglement Purification using Spatial Entanglement,
Phys. Rev. Lett. \textbf{89}, 257901 (2002).


\bibitem{JWPannature2003} J. W. Pan, S. Gasparoni, R. Ursin, G. Weihs, and A. Zeilinger,
Experimental entanglement purification of arbitrary unknown states,
Nature (London) \textbf{423}, 417 (2003).



%\bibitem{WDurPRL2003} W. D\"{u}r, H. Aschauer, and H. J. Briegel,
%Multiparticle Entanglement Purification for Graph States,
%Phys. Rev. Lett. \textbf{91}, 107903 (2003).



%
%\bibitem{YWCheongPRA2007}Y. W. Cheong, S. W. Lee, J. Lee, and H. H. Lee,
%Entanglement purification for high-dimensional multipartite systems,
%Phys. Rev. A \textbf{76}, 042314 (2007).





\bibitem{YBShengPRA2008} Y. B. Sheng, F. G. Deng, and H. Y. Zhou,
Efficient polarization-entanglement purification based on parametric
down-conversion sources with cross-Kerr nonlinearity, Phys. Rev. A \textbf{77}, 042308 (2008).


\bibitem{EPP2} Y. B. Sheng and F. G. Deng, Deterministic entanglement
purification and complete nonlocal Bell-state analysis with hyperentangledment,
Phys. Rev. A \textbf{81}, 032307 (2010).


\bibitem{EPP3} Y. B. Sheng and F. G. Deng,
One-step deterministic polarization-entanglement purification using
spatial entanglement, Phys. Rev. A \textbf{82}, 044305 (2010).



\bibitem{EPP4}X. H. Li, Deterministic polarization-entanglement purification
using spatial entanglement, Phys. Rev. A \textbf{82}, 044304 (2010).


\bibitem{dengonestep} F. G. Deng, One-step error correction for multipartite polarization
entanglement, Phys. Rev. A \textbf{83}, 062316 (2011).


%\bibitem{BCRenLPL2013} B. C. Ren and F. G. Deng, Hyperentanglement purification
%and concentration assisted by diamond NV centers inside photoic
%crystal cavities, Laser Phys. Lett. 10, 115201 (2013).


\bibitem{EPP5} B. C. Ren, F. F. Du, and F. G. Deng,
Two-step hyperentanglement purification with the
quantum-state-joining method, Phys. Rev. A \textbf{90}, 052309
(2014).


\bibitem{GYWangPRA2016} G. Y. Wang, Q. Liu, and F. G. Deng,
Hyperentanglement purification for two-photon six-qubit quantum
systems, Phys. Rev. A \textbf{94}, 032319 (2016).


\bibitem{CWangECPPRA11} C. Wang, Y. Zhang, and G. S. Jin,
Entanglement purification and concentration of electron-spin
entangled states using quantum-dot spins in optical microcavities,
Phys. Rev. A \textbf{84}, 032307 (2011).


\bibitem{EPPadd1} Y. B. Sheng and L. Zhou,
Deterministic polarization entanglement purification using time-bin entanglement,
Laser Phys. Lett. \textbf{11}, 085203 (2014).



\bibitem{EPPadd2} Y. B. Sheng and L. Zhou,
Deterministic entanglement distillation for secure double-server blind quantum computation,
Sci. Rep. \textbf{5}, 7815 (2015).





\bibitem{RFWernerPRA1989} R. F. Werner,
Quantum states with Einstein-Podolsky-Rosen correlations admitting a
hidden-variable model, Phys. Rev. A \textbf{40}, 4277 (1989).





\bibitem{ABlais} A. Blais, R. S. Huang, A. Wallraff, S. M. Girvin, and R. J. Schoelkopf,
Cavity quantum electrodynamics for superconducting electrical circuits:
An architecture for quantum computation, Phys. Rev. A \textbf{69}, 062320 (2004).


\bibitem{AWallraff} A. Wallraff, D. I. Schuster, A. Blais, L. Frunzio,
R. S. Huang, J. Majer, S. Kumar, S. M. Girvin, and R. J. Schoelkopf,
Strong coupling of a single photon to a superconducting qubit using
circuit quantum electrodynamics, Nature (London) \textbf{431}, 162 (2004).




\bibitem{ABlais2} A. Blais, J. Gambetta, A. Wallraff, D. I. Schuster,
S. M. Girvin, M. H. Devoret, and R. J. Schoelkopf, Quantum-information
processing with circuit quantum electrodynamics, Phys. Rev. A \textbf{75}, 032329 (2007).


\bibitem{DiCarlo} L. DiCarlo, J. M. Chow, J. M. Gambetta, Lev S. Bishop,
B. R. Johnson, D. I. Schuster, J. Majer, A. Blais, L. Frunzio, S. M. Girvin,
and  R. J. Schoelkopf, Demonstration of two-qubit algorithms with
a superconducting quantum processor, Nature (London) \textbf{460}, 240 (2009).


\bibitem{LongcircuitPRA} Y. Cao, W. Y. Huo, Q. Ai, and G. L. Long,
Theory of degenerate three-wave mixing using circuit QED in solid-state circuits,
Phys. Rev. A \textbf{84}, 053846 (2011).



\bibitem{Wangsuperconducting} H. Wang, M. Mariantoni, R. C. Bialczak,
M. Lenander, E. Lucero, M. Neeley, A. D. O'Connell, D. Sank, M. Weides,
J. Wenner, T. Yamamoto, Y. Yin, J. Zhao, J. M. Martinis, and A. N. Cleland,
Deterministic entanglement of photons in two superconducting microwave resonators,
Phys. Rev. Lett. \textbf{106}, 060401 (2011).


\bibitem{circuitTianlPRL} Y. Hu and L. Tian,
Deterministic generation of entangled photons in superconducting resonator arrays,
Phys. Rev. Lett. \textbf{106}, 257002 (2011).


\bibitem{3q} A. Fedorov, L. Steffen, M. Baur, M. P. da Silva, and A. Wallraff,
Implementation of a Toffoli gate with superconducting circuits,
Nature (London) \textbf{481}, 170 (2012).


\bibitem{3q1} M. D. Reed, L. DiCarlo, S. E. Nigg, L. Sun, L. Frunzio,
S. M. Girvin, and R. J. Schoelkopf, Realization of three-qubit
quantum error correction with superconducting circuits, Nature  (London) \textbf{482}, 382 (2012).


\bibitem{RVijaynature} R. Vijay, C. Macklin, D. H. Slichter, S. J. Weber, K. W. Murch, R. Naik, A. N. Korotkov, and I. Siddiqi, Stabilizing Rabi oscillations in a superconducting qubit using quantum feedback, Nature (London) \textbf{490}, 77 (2012).


\bibitem{Frederick1} F. W. Strauch, All-resonant control of superconducting resonators,
Phys. Rev. Lett. \textbf{109}, 210501 (2012).









\bibitem{DISchuster} D. I. Schuster, A. A. Houck, J. A. Schreier, A. Wallraff,
J. M. Gambetta, A. Blais, L. Frunzio, J. Majer, B. Johnson, M. H. Devoret,
S. M. Girvin, and R. J. Schoelkopf, Resolving photon number states in a superconducting circuit,
Nature (London) \textbf{445}, 515 (2007).


\bibitem{AAHouck} A. A. Houck, D. I. Schuster, J. Gambetta, J. A. Schreier,
B. R. Johnson, J. M. Chow, L. Frunzio, J. Majer, M. H. Devoret, S. M. Girvin,
and R. J. Schoelkopf, Generating single microwave photons in a circuit,
Nature (London) \textbf{449}, 328 (2007).


\bibitem{JMajer} J. Majer, J. M. Chow, J. M. Gambetta, J. Koch,
B. R. Johnson, J. A. Schreier, L. Frunzio, D. I. Schuster,
A. A. Houck, A. Wallraff, A. Blais, M. H. Devoret, S. M. Girvin,
and R. J. Schoelkopf, Coupling superconducting qubits via a cavity bus,
Nature (London) \textbf{449}, 443 (2007).


\bibitem{MHofheinz} M. Hofheinz, E. M. Weig, M. Ansmann, R. C. Bialczak,
E. Lucero, M. Neeley, A. D. O'Connel, H. Wang, J. M. Martinis,
and A. N. Cleland, Generation of Fock states in a superconducting quantum circuit,
Nature (London) \textbf{454}, 310 (2008).


\bibitem{BRJohnson} B. R. Johnson, M. D. Reed, A. A. Houck,
D. I. Schuster, Lev S. Bishop, E. Ginossar, J. M. Gambetta,
L. DiCarlo, L. Frunzio, S. M. Girvin and R. J. Schoelkopf,
Quantum non-demolition detection of single microwave photons in a circuit,
Nat. Phys. \textbf{6}, 663 (2010).



\bibitem{HuaMPRA} M. Hua,  M. J. Tao, and F. G. Deng,
Universal quantum gates on microwave photons assisted by circuit quantum electrodynamics,
Phys. Rev. A \textbf{90}, 012328 (2014).


\bibitem{Narlaprx2016} A. Narla, S. Shankar, M. Hatridge, Z. Leghtas, K. M. Sliwa, E. Zalys-Geller, S. O. Mundhada, W. Pfaff, L. Frunzio, R. J. Schoelkopf, and M. H. Devoret, Robust Concurrent Remote Entanglement Between Two Superconducting Qubits, Phys. Rev. X \textbf{6}, 031036 (2016).

\bibitem{SRebic2009} S. Rebi\'{c}, J. Twamley, and G. J. Milburn,
Giant Kerr nonlinearities in circuit quantum electrodynamics,
Phys. Rev. Lett. \textbf{103}, 150503 (2009).


\bibitem{SKumarPRB2010} S. Kumar and D. P. DiVincenzo,
Exploiting Kerr cross nonlinearity in circuit quantum electrodynamics
for nondemolition measurements, Phys. Rev. B \textbf{82}, 014512 (2010).


\bibitem{YHu} Y. Hu, G. Q. Ge, S. Chen, X. F. Yang, and Y. L. Chen,
Cross-Kerr-effect induced by coupled Josephson qubits in circuit quantum electrodynamics,
Phys. Rev. A \textbf{84}, 012329 (2011).


\bibitem{GKirchmair} G. Kirchmair, B. Vlastakis, Z. Leghtas,
S. E. Nigg, H. Paik, E. Ginossar, M. Mirrahimi, L. Frunzio, S. M.
Girvin, and R. J. Schoelkopf. Observation of quantum state collapse
and revival due to the single-photon Kerr effect, Nature (London)
\textbf{495}, 205 (2013).


\bibitem{ICHoi} I. C. Hoi, A. F. Kockum, T. Palomaki, T. M. Stace,
B. Fan, L. Tornberg, S. R. Sathyamoorthy, G. Johansson, P. Delsing,
and C. M. Wilson, Giant cross-Kerr effect for propagating microwaves
induced by an artificial atom, Phys. Rev. Lett. \textbf{111}, 053601 (2013).


\bibitem{ETHolland} E. T. Holland, B. Vlastakis, R. W. Heeres,
M. J. Reagor, U. Vool, Z. Leghtas, L. Frunzio, G. Kirchmair,
M. H. Devoret, M. Mirrahimi and R. J. Schoelkopf,
Single-photon-resolved cross-Kerr interaction for
autonomous stabilization of photon-number states, Phys. Rev. Lett. \textbf{115}, 180501 (2015).




\bibitem{JMHao} J. M. Hao, Y. Yuan, L. X. Ran, T. Jiang,
J. A. Kong, C. T. Chan, and L. Zhou, Manipulating electromagnetic wave
polarizations by anisotropic metamaterials, Phys. Rev. Lett. \textbf{99}, 063908 (2007).


\bibitem{DRSolli1} D. R. Solli, C. F. McCormick, R. Y. Chiao,
and J. M. Hickmann, Photonic crystal polarizers and polarizing beam splitters,
J. Appl. Phys. \textbf{93}, 9429 (2003).


\bibitem{JKoch} J. Koch, T. M. Yu, J. M. Gambetta, A. A. Houck,
D. I. Schuster, J. Majer, A. Blais, M. H. Devoret, S. M. Girvin,
and R. J. Schoelkopf, Charge-insensitive qubit design derived
from the Cooper pair box, Phys. Rev. A \textbf{76}, 042319 (2007).


\bibitem{JSiewert} J. Siewert, R. Fazio, G. M. Palma,
and E. Sciacca, Aspects of qubit dynamics in the presence of leakage,
J. Low Temp. Phys. \textbf{118}, 795 (2000).


\bibitem{AImamoglu1997} A. Imamo\v{g}lu, H. Schmidt, G. Woods,
and M. Deutsch, Strongly Interacting Photons in a Nonlinear Cavity,
Phys. Rev. Lett. \textbf{79}, 1467 (1997).


\bibitem{DFWalls} D. F. Walls and G. J. Milburn,
\emph{Quantum Optics} (Springer-Verlag, Berlin, 1994).


\bibitem{Kevin} K. Lalumi\`{e}re, B. C. Sanders, A. F. van Loo, A. Fedorov,
A. Wallraff, and A. Blais, Input-output theory for waveguide QED with
an ensemble of inhomogeneous atoms, Phys. Rev. A \textbf{88}, 043806 (2013).


\bibitem{KNemotoprl2004} K. Nemoto and W. J. Munro,
Nearly Deterministic Linear Optical Controlled-NOT Gate, Phys. Rev. Lett. \textbf{93}, 250502 (2004).


\bibitem{SDBarrettpra} S. D. Barrett, P. Kok, K. Nemoto, R. G. Beausoleil, W. J. Munro, and T. P. Spiller, Symmetry analyzer for nondestructive Bell-state detection using weak nonlinearities, Phys. Rev. A \textbf{71}, 060302(R) (2005).

\end{thebibliography}
\end{document}